\documentclass[graybox, envcountchap]{svmult}
\usepackage{mathptmx}        
\usepackage{amsmath}
\usepackage{amssymb}
\usepackage{color}
\usepackage{helvet}          
\usepackage{courier}         
\usepackage{dirtree}
\usepackage[english]{babel}
\usepackage{ulem}
\usepackage{makeidx}        
\usepackage{graphicx}        
\usepackage{subfig}
\usepackage{amssymb}

\usepackage{multicol}        
\usepackage[bottom]{footmisc}

\usepackage{hyperref}        
\usepackage{mathrsfs}
\hypersetup{colorlinks=true,urlcolor=blue}
\usepackage[dvipsnames]{xcolor}
\usepackage[misc]{ifsym}

\makeindex             

\begin{document}


\title{Deriving the paradox: original derivation of Hawking radiation}
\author{Mariano Cadoni, Salvatore Mignemi, Mirko Pitzalis and Andrea P. Sanna}
\institute{Mariano Cadoni \at Dipartimento di Fisica, Universit\`a di Cagliari, Cittadella Universitaria, 09042 Monserrato, Italy,\\
I.N.F.N, Sezione di Cagliari, Cittadella Universitaria, 09042 Monserrato, Italy\\
 \email{mariano.cadoni@ca.infn.it}
 \and Salvatore Mignemi \at Dipartimento di Matematica, Universit\`a di Cagliari, Via Ospedale 72, 09124 Cagliary, Italy, \\
 INFN Sezione di Cagliari, Cittadella Universitaria, 09042 Monserrato, Italy\\
 \email{smignemi@unica.it}
\and Mirko Pitzalis \at Dipartimento di Fisica, Universit\`a di Cagliari, Cittadella Universitaria, 09042 Monserrato, Italy,\\
I.N.F.N, Sezione di Cagliari, Cittadella Universitaria, 09042 Monserrato, Italy\\
\email{mirko.pitzalis@ca.infn.it} \and
Andrea P. Sanna (\Letter) \at Dipartimento di Fisica, Universit\`a di Cagliari, Cittadella Universitaria, 09042 Monserrato, Italy,\\
I.N.F.N, Sezione di Cagliari, Cittadella Universitaria, 09042 Monserrato, Italy\\
\email{asanna@dsf.unica.it}
}
%
%
\maketitle

\abstract{
We revisit Hawking's original derivation of the evaporation process in a non-stationary spacetime, presenting it in a clear and pedagogical manner, with a focus on the spherical collapse of a star into a black hole. Our analysis highlights the underlying assumptions in the calculations, clarifying their physical significance, potential implications, and the limitations of this approach.}

\section{Introduction}
\label{sec:Intro}

Hawking's original derivation of black-hole thermal emission is almost fifty years old~\cite{Hawking:1975vcx}. To date it stands as a cornerstone of theoretical physics, combining physical intuition with mathematical rigour, each complementing the other as the derivation grows increasingly complex.

Hawking's paper not only firmly established black-hole thermodynamics on solid physical grounds, elevating it from a mere analogy, but also paved the way for the "golden age" of quantum field theory (QFT) in curved spacetime. It sparked groundbreaking ideas and developments among the physics community, even beyond black-hole physics, ranging from the Fulling-Davies-Unruh effect~\cite{Fulling:1972md,Davies:1974th,Unruh:1976db}, cosmological particle creation~\cite{Parker:1969au,Parker:1971pt,Parker:1976qf,Zeldovich:1970si,Zeldovich:1971mw}, quantum information theory up to black-hole analogues in condensed matter physics~\cite{Page:1993wv,Penington:2019npb,Almheiri:2020cfm,Raju:2020smc,Barcelo:2000tg,Weinfurtner:2010nu,Drori:2018ivu}. Hawking radiation also lies at the root of one of the most challenging open problems in fundamental theoretical physics: the black-hole information problem~\cite{Page:1993wv,Penington:2019npb,Almheiri:2020cfm,Hawking:1976ra,Penington:2019kki,Almheiri:2019qdq}.

The literature now offers several derivations of the Hawking effect, covered here in Chapter $2$. Each of them employs different setups for the background geometry, ranging from collapsing bodies to eternal black holes. Some of them also differ in their characterization of the quantum effects at play (like, for instance, conformal anomalies). Additionally, derivations that address the information problem and concepts like complementarity~\cite{tHooft:1984kcu,tHooft:1990fkf,Susskind:1993if,Lowe:1995ac,Kiem:1995iy}, such as the Firewall proposal~\cite{Almheiri:2012rt}, typically use a set of quantum-field modes that are purely ingoing at the horizon, whereas the original Hawking derivation employs ingoing modes at null past infinity.\\

Here we present a revisited version of the original Hawking derivation.
We closely follow the line of reasoning of the original derivation, modifying it only in some minor technical aspects, aiming at enhancing its pedagogical clarity and simplifying certain details that are not essential for the understanding of the whole subject.

In the following sections, we will first discuss the physical framework (Sect.~\ref{sect:1}), followed by an introduction to the basics of QFT in curved spacetime (Sect.~\ref{sec:2}). Finally, we will present the derivation of the Hawking effect (Sect.~\ref{sec:3}).

\section{Hawking radiation: Physical framework }
\label{sect:1}

The strongest argument in favor of black-hole thermal emission is provided by black-hole thermodynamics. In the stationary uncharged case, the mass $M$, the horizon area $A$ and the angular momentum $J$ satisfy the differential relation~\cite{Bardeen:1973gs}
\begin{equation}
\label{firstlaw}
\text{d} M= \frac{{\cal{K}}}{8\pi}  \text{d}A + \Omega \text{d}J\, ,
\end{equation}
where ${\cal K}$ is the surface gravity at the event horizon and $\Omega$ the black-hole angular velocity. This expression is analogous to the first law of thermodynamics if one identifies $A$ as the black-hole entropy (as first conjectured by Bekenstein~\cite{Bekenstein:1972tm,Bekenstein:1973ur}) and ${\cal K}/2\pi$ as the temperature \footnote{Throughout the chapter, we will adopt units in which $G = \hslash =  c = k_\text{B}= 1$}.
Furthermore, as argued by Jacobson in~\cite{Jacobson:1995ak,Jacobson:1995ab}, this is not merely an analogy: a black hole must genuinely behave as a thermodynamic system. If this is true, consistency with the basic principles of thermodynamics and statistical mechanics requires a black hole to be capable of both emitting and absorbing particles. The primary conceptual challenge in this reasoning, as opposed to a standard black body, lies in the lack of a microscopic description of black holes. Hawking's striking physical intuition was that, being gravity linked to curvature, particle creation by a black hole can be understood as a consequence of the dynamics of quantum fields  in a curved spacetime.
Qualitatively, particle creation in a region of spacetime with curvature of the order of $R$ can be understood as resulting from the uncertainty in the particle number operator $a^{\dagger} a$  for modes with wavelengths larger than $\sim R^{-1/2}$. This arises because defining positive frequency modes requires a local inertial coordinate system, which can only be set up within a region of size $\sim R^{-1/2}$. The uncertainty in the particle number operator leads to an uncertainty in the local energy density of order $R^2$. This uncertainty, therefore, can be thought of as corresponding to the local energy density of particles created by the gravitational field. As long as $R^{-1/2}$ remains much larger than the Planck length $\ell_\text{P}$, Einstein's equations imply that the corresponding uncertainty in the spacetime curvature due to this energy density uncertainty can be safely neglected. Consequently, the semiclassical description  provides an ideal framework for discussing particle creation by macroscopic black holes. In this framework, gravity is treated classically, while the matter fields are quantized on the curved black-hole background. Following the same reasoning, the gravitational background of a large black hole can be treated as fixed, allowing backreaction effects on the geometry to be neglected as a first approximation.

A consequence of particle emission is the decrease in the black-hole mass and in the area of the event horizon \footnote{Black-hole evaporation represents a violation of the area theorem~\cite{Hawking:1971tu}, which implies that the horizon area can never decrease in any dynamical process. It is important to stress, however, that the theorem rests on the validity of the weak energy condition, which is violated during quantum thermal emission. This violation can be understood as a consequence of the indeterminacy of particle number and energy density in a curved spacetime mentioned above.}. This process can be understood as arising from a negative energy flux across the event horizon, balancing the positive energy flux emitted to infinity. Heuristically, the negative energy flux is due to the tunneling of negative energy particles created from the continuous spontaneous creation of virtual particle-antiparticle pairs around the black hole. These particles emerge from a classically forbidden region, tunnel through the event horizon and enter the black hole, where the time-translation Killing vector becomes spacelike. This allows negative energy particles to exist as real particles with a timelike momentum vector having negative energy relative to infinity, as measured by the time translation Killing vector. The positive energy partner of the virtual pair can escape to infinity, where it contributes to the thermal emission. The probability of the negative energy particle tunneling through the event horizon is governed by the surface gravity $\cal{K}$, as this quantity measures how fast the Killing vector becomes spacelike. Virtual particle pairs created with wavelength $\lambda$ separate temporarily to a distance $\sim\lambda$. For $\lambda$ of the order of the black-hole mass, strong tidal forces prevent re-annihilation, allowing the black hole to radiate quanta with wavelengths of order of $M$. The thermal emission carries away energy, causing the black hole to lose mass.

However, in most cases, particle creation is locally negligible. The black-hole temperature is roughly $T \sim 10^{-6} \, \left(M_\odot/M \right) \, \text{K}$, meaning that for solar-mass black holes, it is much lower than that of the Cosmic Microwave Background. As a result, these black holes absorb more radiation than they emit. It may nonetheless lead to significant global effects when considered over time scales comparable to the lifetime of the Universe. For smaller black holes, for instance, such as those potentially formed in the early Universe~\cite{Hawking:1971ei,Escriva:2022duf}, the Hawking process can drive their complete evaporation.\\

Another remarkable aspect of the Hawking effect, which can be inferred from general principles, is the universality of the radiation flux. Surprisingly, it is independent of the details of the gravitational collapse and the form of the quantum matter fields. This property can be logically deduced from the no-hair theorems~\cite{Israel:1967wq,Israel:1967za,Carter:1971zc}. This indicates that the key factor is the event-horizon structure. This universality, however, holds strictly only when greybody factors are neglected, as we will assume in the following sections.

This universality can also be leveraged to simplify calculations by considering the simplest setup: massless scalar quantum fields in the simplest spherically-symmetric collapsing geometry, the Vaidya spacetime. To describe the semiclassical dynamics of quantum fields, it is necessary to define a complete set of purely ingoing and outgoing modes, each containing only positive frequencies relative to the time coordinate used by the corresponding observer. There are several methods to achieve this. In the next section, we will adopt the approach originally used by Hawking~\cite{Hawking:1975vcx}. We define:
\begin{itemize}
\item[$(a) $] \quad  a  complete set of purely ingoing modes $\phi^{in}$, containing only positive frequencies with respect to the affine parameter on past null infinity $\cal{I}^-$. These are completely determined by their Cauchy data on $\cal{I}^-$;
\item[$(b)\, $] \quad a complete set of modes constructed from purely outgoing $\phi^{out}$  and internal modes $\phi^{int}$. The former contains only positive frequencies relative to the affine parameter at future null infinity $\cal{I}^+$ and has zero Cauchy data on $\cal{I}^-$. The latter has Cauchy data on the event horizon, but zero data on both $\cal{I}^-$ and $\cal{I}^+$.
\end{itemize}
As long as one is concerned with calculating the outgoing flux of Hawking particles at late times, the $\phi^{int}$ modes seem to be completely irrelevant. Particle creation is a global process not localized on the event horizon. The key to understanding this lies in the relationship between the ``in'' and ``out'' modes, governed by a Bogoliubov transformation. This transformation underpins the entire framework: from Hawking radiation to the information paradox.

Furthermore, as long as the flux of Hawking particles at $\cal{I}^+$ is considered, a different set of modes can be used, i.e., $\phi^{out}$ and $\phi^{in}_h$ instead of $\phi^{in}$ and $\phi^{int}$. The modes $\phi^{in}_h$ have positive frequency relative to a free-falling observer near the event horizon. This setup is commonly adopted in recent formulations addressing the information problem, like those leading to the Firewall hypothesis~\cite{Almheiri:2012rt}. This approach will not be employed in the derivation contained in the following sections. \\

Hawking's derivation is far from being devoid of conceptual issues, as it is known to be plagued by the so-called transplanckian problem~\cite{Jacobson:1991gr,Jacobson:1993hn,Massar:1996tx,Jacobson:1996zs}. As detailed in Sect.~\ref{sec:3}, Hawking's results involve integrating outgoing modes through the collapsing body. This is unphysical due to blue-shift effects. Nevertheless, Hawking recognized that the essential element of his derivation was the horizon structure, implying that the results should be independent of ultraviolet (UV) gravitational physics.

In the modern framework based on the modes $\phi^{out}$ and $\phi^{in}_h$, this independence from UV physics is rephrased in terms of the adiabatic principle. By considering a free-falling observer near the horizon and foliating the near-horizon geometry with smooth slices, it becomes clear that the geometry evolves slowly compared to the blue-shifted frequency of  the Hawking modes at $\cal{I}^+$. Consequently, according to the adiabatic principle, these modes must remain in their ground state relative to the infalling observer frame. Even though, to date, a complete description of UV gravitational physics remains elusive, as long as the adiabatic and equivalence principles hold, the extreme blue-shift remains irrelevant, and all relevant information is encoded in the event-horizon structure. \\

In the following sections, we will go through the derivation of the Hawking effect using the setup elucidated in this section. For the technical computation, we will mainly follow Ref.~\cite{Fabbri:2005mw}.

\section{Quantization in curved spacetime}
\label{sec:2}

In a curved spacetime, unlike in the flat Minkowski one, Poincar\'e symmetry is no longer preserved. As a consequence, one cannot globally define field modes with positive frequency across the entire spacetime, because in curved spacetime they lose their invariant meaning, as also the notion of vacuum or particle number.  Two local free-falling observers at two different points in spacetime will each employ distinct timelike coordinates, leading to two different notions of positive frequency modes.
Thus, the standard quantization procedure must be extended to accommodate curved geometries. The minimal generalization is as follows
\begin{itemize}
\item We consider the dynamics of matter fields in a fixed curved gravitational background, focusing specifically on massless scalar fields $\phi$. At the classical level, $\phi$ satisfies the standard wave equation, with the Minkowski metric replaced by a classical spacetime metric $g_{\mu\nu}$, solution of Einstein's equations. Therefore, the Klein-Gordon (KG) equation is generalized as usual by simply replacing partial derivatives with covariant derivatives
\begin{equation}
\Box \phi = g^{\mu\nu}\nabla_\mu \nabla_\nu \phi = 0\, ;
\label{KGequationgeneral}
\end{equation}
\item Assuming the spacetime to be globally hyperbolic ensures the existence of an initial data Cauchy hypersurface $\Sigma$, on which the KG inner product can be generalized and computed as follows
\begin{equation}
(\phi_1\, , \phi_2) = -\text{i} \int_\Sigma \text{d}\Sigma^\mu \left(\phi_1 \partial_\mu \phi_2^{\ast} -  \phi_2^{\ast}\partial_\mu \phi_1\right)\, ,
\label{KGproduct}
\end{equation}
where $\text{d}\Sigma^\mu = \text{d}\Sigma \, n^\mu$, with $\text{d}\Sigma$ representing the spacelike volume element, while $n^\mu$ the future-directed unit normal vector to $\Sigma$.

A key property of this inner product is that it is \textit{independent} of the choice of $\Sigma$. To see this, consider two hypersurfaces $\Sigma$ and $\Sigma'$, enclosing a four-dimensional volume $S$. The difference between the inner product computed on $\Sigma$ and on $\Sigma'$ can be expressed as a volume integral over $S$, thanks to Gauss' law. Using the KG equation \eqref{KGequationgeneral}, one finds $(\phi_1\, , \phi_2)_\Sigma - (\phi_1\, , \phi_2)_{\Sigma'}=0$, confirming the assertion.
\end{itemize}
In flat spacetime, the standard quantization procedure involves decomposing the field $\phi$, solution of the KG equation $\eta^{\mu\nu} \partial_\mu \partial_\nu \phi=0$, into positive and negative frequency modes
\begin{equation}
\phi= \sum_i \left(\phi_i \,  a_i + \phi_i^\ast \, a_i^\dagger \right)\, ,
\end{equation}
where the $\left\{\phi_i \right\}$ form a complete orthonormal set of solutions to the KG equation. Here, $a_i$ and $a_i^\dagger$ represent the annihilation and creation operators, respectively, for particles in the i-th state. The vacuum state is uniquely defined as the state for which $a_i|0\rangle = 0$ for all $i$.

In curved spacetime, however, the splitting between positive and negative frequency modes is not unique, leading to the absence of a single, well-defined vacuum state. Different choices of positive frequency modes result in different vacuum states and hence a different  Fock space. Indeed, since generically there will not be any timelike Killing vector field, one will not in general be able to find solutions to the wave equation that separate into time-dependent and space-dependent factors, and correspondingly will not be able to classify modes into positive- or negative- frequency ones. Nevertheless, in stationary spacetime, we can still select a natural set of positive frequency solutions. In this case, a timelike Killing vector field $\xi^\mu$ exists, leaving the metric invariant under infinitesimal transformations expressed as $\delta_{\xi}g_{\alpha \beta} = 0$. This allows for a global definition of positive frequency modes as those satisfying
\begin{equation}
\xi^\mu \nabla_\mu \phi_j = -\text{i} \omega_j \phi_j \, , \qquad \omega_j >0 \, .
\end{equation}
In more general settings, where the spacetime is non-stationary (for instance during gravitational collapse) or when the timelike Killing vector field is not globally defined (e.g., in the presence of an event horizon), it is no longer possible to define globally a set of positive frequency modes. Even in a general case, one can still have a natural choice of positive frequency modes for those spacetimes possessing asymptotic stationary regions in the past and in the future, commonly referred to as ``in'' and ``out'' regions, respectively. There, one can construct orthonormal sets of modes, which are exact solutions of the wave equation, with positive frequency with respect to the inertial time in the past $\left\{\phi_i^{in} \right\}$ and in the future $\left\{\phi_i^{out} \right\}$.

A key consequence of not having a unique, globally-defined vacuum state is that, if quantum fields start in the vacuum state as observed by an inertial observer in the "in" region $|in\rangle$, the changing, non-stationary gravitational field causes these fields to no longer appear in the vacuum state for an inertial observer in the "out" region at later times. This phenomenon can be precisely quantified by resorting to the Bogoliubov transformations.

\subsection{Bogoliubov transformations}
\label{subsec:Bogotransfo}

Let us expand the scalar field $\phi$ in the basis of the positive frequency solutions $\left\{\phi_i^{in}\right\}$ in the "in" stationary region
\begin{equation}
 \phi = \sum_i \left[a_i^{in} \phi_i^{in} + a^{\dagger in}_i \phi_i^{in \ast} \right]\, ,
\label{uinexpansion}
\end{equation}
while the same expansion can be performed in the "out" stationary region, in terms of the basis $\left\{\phi_i^{out}\right\}$
\begin{equation}
\phi= \sum_i \left[a_i^{out} \phi_i^{out} + a_i^{\dagger out} \phi_i^{out \ast} \right]\, .
\label{uoutexpansion}
\end{equation}
As discussed in Sect.~\ref{sect:1}, the ``out'' region represents a Cauchy hypersurface only if we include the event horizon. In other words, a complete set of modes in the ``out'' region is represented by purely outgoing modes $\phi^{out}$ and modes in the interior of the black hole $\phi^{int}$. However, we will neglect the latter, as they do not alter particle production in the ``out'' region.

The modes satisfy the relations, both valid for ``in'' and ``out'' modes (we omit the superscripts)
\begin{equation}
\left(\phi_i, \, \phi_j \right) = -(\phi_i^\ast, \, \phi_j^\ast) = \delta_{ij}\, , \qquad (\phi_i, \, \phi_j^\ast) = 0\, ,
\label{normalizationmodes}
\end{equation}
while the creation and annihilation operators satisfy the commutation relations (again we drop the superscripts ``in'' and ``out'')
\begin{equation}
\left[a_i, \, a_j^\dagger \right] = \delta_{ij} \, , \qquad \left[a_i, \, a_j \right] = \left[a_i^\dagger, \, a_j^\dagger \right] = 0\, .
\label{commutationelationsoperators}
\end{equation}
Since both sets of modes are complete, it is possible to expand one set in the basis of the other. However, it is not guaranteed that positive frequency solutions of one set can be expanded solely in terms of the pure positive frequency solutions of the other. As a result, the expansion will generally involve a mixture of both positive and negative modes, with the expansion coefficients given by the Bogoliubov matrices $\alpha$ and $\beta$, namely
\begin{equation}
\phi_j^{out} = \sum_i \left(\alpha_{ji} \, \phi_i^{in} + \beta_{ji}\,  \phi_i^{in \ast} \right)\, .
\label{Bogoliubovmatrices}
\end{equation}
Using the orthonormal relations \eqref{normalizationmodes}, we can invert the above to write
\begin{equation}
\alpha_{ij} = \left(\phi_i^{out}, \, \phi_j^{in} \right)\, , \qquad \beta_{ij} = -\left(\phi_i^{out}, \, \phi_j^{in \ast} \right)\, .
\label{alphabetaBogo}
\end{equation}
Equation \eqref{normalizationmodes} can also be used to write the following relations between the Bogoliubov coefficients, which will be essential in deriving the spectrum of Hawking radiation
\begin{equation}
\begin{split}
& \sum_k \left(\alpha_{ik} \alpha_{jk}^\ast - \beta_{ik} \beta_{jk}^\ast \right) = \delta_{ij}\, ;\\
&\sum_k \left(\alpha_{ik} \beta_{jk} - \beta_{ik} \alpha_{jk} \right) = 0\, .
\end{split}
\label{Bogocoeffrelations}
\end{equation}
We can also invert eq. \eqref{Bogoliubovmatrices} and write
\begin{equation}
\phi_i^{in} = \sum_j \left(\alpha_{ji}^\ast \, \phi_j^{out} - \beta_{ji} \, \phi_j^{out \ast} \right)\, .
\end{equation}

Using the previous relations, together with the fact that, from eq. \eqref{uinexpansion} and eq. \eqref{uoutexpansion}, we also have $a_i^{in} = (\phi, \, \phi_i^{in})$ and $a_i^{out} = (\phi, \, \phi_i^{out})$, respectively, we can expand one of the two sets of creation and annihilation operators in terms of the other
\begin{equation}
\begin{split}
&a_i^{in} = \sum_j \left(\alpha_{ji} \, a_j^{out} + \beta_{ji}^\ast \, a_j^{\dagger out} \right)\, ;\\
&a_i^{out} = \sum_j \left(\alpha_{ij}^\ast \, a_j^{in} - \beta^\ast_{ij} \, a_j^{\dagger in} \right)\, .
\end{split}
\end{equation}
As always, the ``in'' and ``out'' vacuum states are defined as $a_i^{in}|in\rangle = 0$ and $a_i^{out}|out\rangle = 0$ for all $i$, respectively. We can exploit this to compute the expectation value of the ``out'' particle number operator for the i-th mode
\begin{equation}
N_i^{out} \equiv a_i^{\dagger out} a_i^{out}\, ,
\end{equation}
in the state $|in\rangle$
\begin{equation}
\begin{split}
\langle in |N_i^{out}|in \rangle & = \langle in |a_i^{\dagger out} a_i^{out} |in\rangle\\
&= \langle in |\sum_j \left(-\beta_{ij} a_j^{in} \right)\sum_k \left(-\beta_{ik}^\ast \, a_k^{\dagger in} \right)|in\rangle\\
& = \sum_j \left|\beta_{ij}\right|^2\, .
\end{split}
\label{expectationvalueN}
\end{equation}
Thus, if the $\beta_{ij}$ coefficients are non-zero, the particle content of the $|in\rangle$ vacuum state, with respect to the ``out'' Fock space, is non-trivial. Thus, what appears as an empty vacuum from one perspective is seen as teeming with particles from another. This exactly corresponds to particle creation in the "out" region from the vacuum in the "in" region.
Conversely, if all $\beta_{ij}$ vanish, we have, from eq.~\eqref{Bogocoeffrelations} that $\sum_k \alpha_{ik} \alpha^\ast_{jk} = \delta_{ij}$, meaning that the positive frequency mode bases $\left\{u_i^{in} \right\}$ and $\left\{u^{out}_i \right\}$ are related by a unitary transformation. In this case, the definition of the vacuum remains unchanged, i.e., $|in\rangle = |out \rangle$. \\
The main result is that the $\beta_{ij}$ coefficients are the essential ingredient to determine the number of particles created  by the gravitational field during the gravitational collapse and detected by an observer in the asymptotic region.
As we will see, the Bogoliubov coefficients have a simple and \textit{universal} asymptotic form, depending only on the surface gravity of the newly formed black hole.


\section{Hawking radiation from a collapsing body }
\label{sec:3}
\subsection {The Vaidya spacetime}

\begin{figure}[!t]
\centering
\includegraphics[scale=0.5]{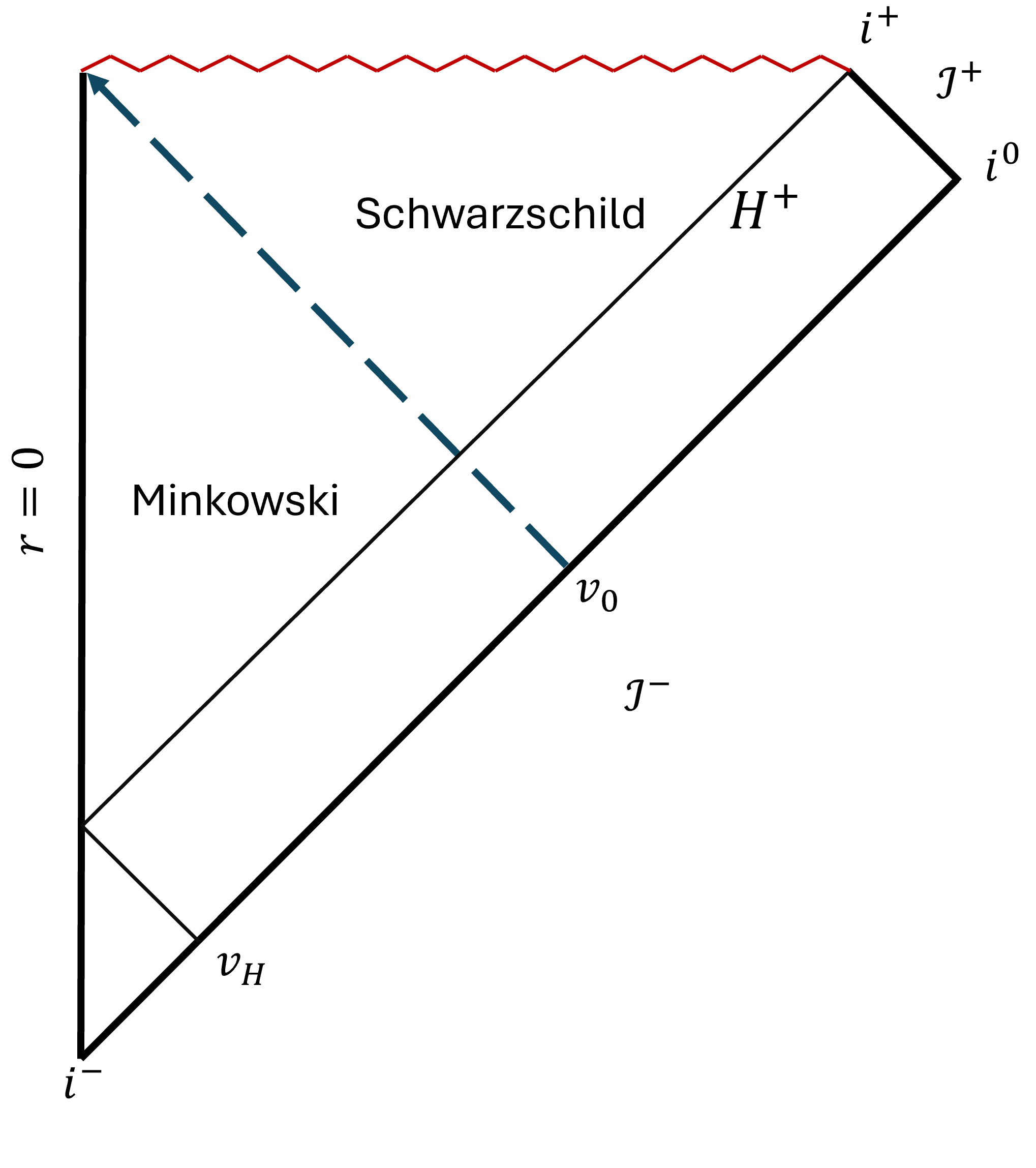}
%
%
\caption{Penrose diagram of the Vaidya spacetime. A Schwarzschild solution at late times is jointed with a Minkowski one at early times along the shock-wave at advanced time $v = v_0$ of the ingoing flux of matter (blue dashed arrow).}
\label{fig:Vaidya}       
\end{figure}

To understand the origin of particle creation, it is essential to consider not only the quasi-stationary final state of the black hole, but also the time-dependent formation phase. The Vaidya spacetime provides the simplest solution of Einstein's equations describing a spherically-symmetric collapse leading to black-hole formation. In Eddington-Finkelstein coordinates, the metric is given by
\begin{equation}
\text{d}s^2 = -\left(1-\frac{2M(v)}{r} \right)\text{d}v^2 + 2 \text{d}v \text{d}r + r^2 \text{d}\Omega^2\, , \quad \text{d}\Omega^2 = \text{d}\theta^2 + \sin^2 \theta \text{d}\varphi^2\, ,
\end{equation}
where $v$ is the advanced time. When $M(v)  = \text{constant} \equiv M\neq 0$, we recover the static Schwarzschild spacetime, whereas when $M(v)=0$ the metric reduces to the flat Minkowski spacetime. Instead, if the mass is solely a function of $v$, the stress-energy tensor sourcing the solution takes the form
\begin{equation}
T_{vv} = \frac{L(v)}{4\pi r^2} \, , \qquad \frac{\text{d}M}{\text{d}v} = L(v) \, ,
\end{equation}
describing a purely ingoing radial flux of radiation and representing the simplest case of gravitational collapse. In more realistic scenarios, there would also be an outgoing flux as well as non-spherical deformations. We assume, additionally, that the influx radiation is turned on at some finite advanced time $v_i$ and turned off at $v_f$. In this way, we can divide the spacetime into three regions
\begin{enumerate}
\item For $v < v_i$, $M(v)=0$ and we have flat Minkowski spacetime;
\item For $v_i < v < v_f$, $M(v)$ grows and we have the collapse region;
\item For $v > v_f$, $M(v)=M$ and we have the final Schwarzschild black hole.
\end{enumerate}
The original analysis considers the complete gravitational collapse of a spherical distribution of matter. As noted by Hawking, however, only the regions outside the collapsing matter in the asymptotic future are relevant. While the exact form of the Bogoliubov coefficients will depend somewhat on the details of the gravitational collapse, their asymptotic form is completely determined by the surface gravity of the black hole. Instead, the modes of any quantum field propagating through the interior of the collapsing body will be severely disrupted; thus, even though some particle creation will depend on the details of the collapse, these particles will disperse. In conclusion, only the first and last regions above are important and, at late retarded times at $\cal{I}^+$, there will be a steady flux of particles fully determined by the asymptotic form of the Bogoliubov coefficients. The discussion above, therefore, allows one to make a further simplifying assumption: the region $2$ can be freely narrowed down to an idealized single null surface. This is equivalent to treating the ingoing flux as a shock-wave at some $v = v_0$, namely
\begin{equation}
L(v) = M \delta(v-v_0) \, , \qquad M(v) = M \theta(v-v_0)\, .
\end{equation}
Thus, the full solution is obtained by jointing the Minkowski and Schwarzschild spacetimes across the shock-wave. The Minkowski region corresponds to the ``in'' region, while the Schwarzschild one represents the ``out'' region (see Fig.~\ref{fig:Vaidya}). Introducing a shockwave that connects a Schwarzschild spacetime to Minkowski spacetime does not alter the original analysis, but instead provides a clearer definition of the initial conditions.

\subsection{Massless fields and wave equations}

We consider the simplest case of matter fields propagating in the Vaidya spacetime: a massless scalar field, obeying the KG equation \eqref{KGequationgeneral}. Given that the background geometry is spherically symmetric, the mode solution of the wave equation~\eqref{KGequationgeneral} can be decomposed into time, radial and angular components, with the latter expressed by spherical harmonics $\mathcal{Y}_{lm} (\theta, \, \varphi)$
\begin{equation}
\phi(t, r, \theta, \varphi) = \sum_{l, m} \frac{\phi_l(t, r)}{r} \, \mathcal{Y}_{lm} (\theta, \, \varphi)\, .
\end{equation}
Neglecting the angular part, the problem reduces to a two-dimensional one in $t$ and $r$ coordinates. Since we have two jointed regions, we will have two corresponding equations
\begin{itemize}
\item In the ``in'' Minkowski region $v < v_0$, we have
\begin{equation}
\left[-\partial_t^2 + \partial_r^2 -\frac{l(l+1)}{r^2}\right]\phi_l(t, r) = 0\, ;
\label{KGin}
\end{equation}
\item In the ``out'' Schwarzschild region $v > v_0$, we instead have
\begin{equation}
\left[-\partial_t^2 + \partial_{r_\ast}^2 -\mathcal{V}_l(r) \right]\phi_l(t, r) = 0\, ,
\label{KGout}
\end{equation}
where $r_\ast$ is the tortoise coordinate
\begin{equation}
r_\ast = \int \frac{\text{d}r}{1-\frac{2M}{r}} = r + 2M \ln \left|\frac{r}{2M}-1 \right|\, ,
\label{tortoisecoordSchw}
\end{equation}
while $\mathcal{V}_l(r)$ is the effective potential
\begin{equation}
\mathcal{V}_l(r) \equiv \left(1-\frac{2M}{r} \right)\left[\frac{l(l+1)}{r^2}+\frac{2M}{r^3} \right]\, .
\label{effpotential}
\end{equation}
\end{itemize}
Note that the potential vanishes both at the event horizon $r = 2M$ ($r_\ast \to -\infty$) and at spatial infinity $r \to \infty$ ($r_\ast \to +\infty$). Since the relevant physics occurs near the horizon, we introduce a further simplification and neglect the potential everywhere. This approximation is justified within the geometric optics approximation and is particularly well-suited for the ``s-wave'' component, i.e., the one with $l = 0$, as this mode is the least affected by the potential.

Thus, within this approximation, we deal with free KG equations in both regions, the only difference being the form of the radial coordinate. We are, therefore, left with the following two equations
\begin{itemize}
\item For $v < v_0$
\begin{equation}
\left(-\partial_t^2 + \partial_r^2 \right)\phi(t, r) = 0\, ,
\end{equation}
with the boundary condition $\phi(t, r = 0) = 0$ to ensure regularity (notice that we dropped the subscript $l$ in $\phi_l$ since we are considering only the s-wave sector).
\item For $v > v_0$
\begin{equation}
\left(-\partial_t^2 + \partial_{r_\ast}^2  \right)\phi(t, r) = 0\, .
\end{equation}
\end{itemize}
We will allow only positive frequency modes. Thus, we assume the following harmonic dependence on time
\begin{equation}
\phi(t, r) = e^{-\text{i}\omega t} \phi(r) \, ,
\end{equation}
and the KG equations become
\begin{equation}
\begin{split}
&\frac{\text{d}^2 \phi}{\text{d}r^2} + \omega^2 \phi(r) = 0 \, , \qquad \text{``in'' region}\\
&\frac{\text{d}^2 \phi}{\text{d}r_{\ast}^2} + \omega^2 \phi(r) = 0 \, , \qquad \text{``out'' region}
\end{split}
\end{equation}
The solutions of these equations are better expressed in terms of null coordinates $u = t - r$ and $v = t + r$. Since we have two different spacetimes across the shock-wave, we will have different transformations
\begin{itemize}
\item In the Minwkoski ``in'' region, we define $u_{in} = t_{in} - r_{in}$ and $v = t_{in} + r_{in}$ and the metric becomes
\begin{equation}
\text{d}s_{in}^2 = -\text{d}u_{in} \text{d} v + r^2_{in} \text{d}\Omega^2\, .
\end{equation}
\item In the Schwarzschild ``out'' region, we instead define $u_{out} = t_{out}-r_{out}^\ast$ and $v = t_{out} + r_{out}^\ast$, and the metric reads
\begin{equation}
\text{d}s_{out}^2 = -\left(1-\frac{2M}{r_{out}} \right) \text{d}u_{out} dv + r_{out}^2 \text{d}\Omega^2\, .
\end{equation}
\end{itemize}
The solutions of the wave equations will consist of a combination of ingoing and outgoing waves. In both cases, the ingoing waves will have a time dependence of the form $\sim e^{-\text{i}\omega v}$. The outgoing waves, however, will behave as $\sim e^{-\text{i}\omega u_{in}}$ in the Minkowski region, and as $\sim e^{-\text{i}\omega u_{out}}$ in the Schwarzschild one.

\subsection{Ingoing and outgoing modes}
We first define the ``in'' Fock space associated with the natural time parameter at $\mathscr{I}^-$. The complete set of positive frequency modes are
\begin{equation}
\phi_\omega^{in} = \frac{1}{4\pi \sqrt{\omega}} \frac{e^{-\text{i} \omega v}}{r}\, .
\label{positivemodesinregion}
\end{equation}
These  modes are completely determined by their Cauchy data on $\cal{I}^-$.

It is straightforward to show that they satisfy the normalization condition
\begin{equation}
(\phi_\omega^{in}, \, \phi_{\omega'}^{in}) = -\text{i}\int_{\mathscr{I}^-} \text{d}v \, \text{d}\Omega^2 \, r^2 \left(\phi_\omega^{in} \partial_v \phi_{\omega'}^{in \ast} - \phi_{\omega'}^{in \ast} \partial_v \phi_\omega^{in} \right) = \delta(\omega-\omega')\, .
\end{equation}
Then, we can construct the Fock space at $\mathscr{I}^+$, associated, instead, to the natural time parameter at $\mathscr{I}^+$, whose   complete set of positive frequency modes is
\begin{equation}
\phi_\omega^{out} = \frac{1}{4\pi \sqrt{\omega}} \frac{e^{-\text{i}\omega u_{out}}}{r}\, .
\label{positivemodesoutregion}
\end{equation}

These modes have zero Cauchy data on $\cal{I}^-$ and satisfy the normalization condition
\begin{equation}
(\phi_\omega^{out}, \, \phi_{\omega'}^{out}) = -\text{i}\int_{\mathscr{I}^+} \text{d}u_{out} \, \text{d}\Omega^2 \, r^2 \left(\phi_\omega^{out} \partial_{u_{out}} \phi_{\omega'}^{out \ast} - \phi_{\omega'}^{out \ast} \partial_{u_{out}} \phi_\omega^{out} \right) = \delta(\omega-\omega')\, .
\end{equation}
$\mathscr{I}^+$ is not a proper Cauchy surface in the ``out'' region, but we must add also the future horizon $H^+$. In other words, the modes $\phi_\omega^{out}$ do not form a complete space. To construct the complete ``out'' Fock space, we must add the modes that cross the future horizon $\phi_\omega^{int}$.
As first noted in Sect.~\ref{sect:1}, however, it turns out that the final expression describing particle production at future infinity is insensitive to the specific expression of these modes.

In subsection \ref{subsec:Bogotransfo}, we have seen that the most important ingredient to describe particle production is the Bogoliubov coefficient $\beta$, given by eq.~\eqref{alphabetaBogo}. In this case, we have
\begin{equation}
\begin{split}
\beta_{\omega \omega'} & = -\left(\phi_\omega^{out}, \, \phi_{\omega'}^{in \ast} \right)= \text{i}\int_{\mathscr{I}^-} \text{d} v \, \text{d}\Omega^2 \, r^2 \left(\phi_\omega^{out} \partial_v \phi_{\omega'}^{in}  -  \phi_{\omega'}^{in} \partial_v \phi_\omega^{out}\right)\, .
\label{betaBogoomegaomegaprime}
\end{split}
\end{equation}
Since the scalar product is independent of the choice of the Cauchy surface, for mathematical and physical convenience we choose to evaluate it at $\mathscr{I}^-$ . However, to do so, we must compute the behavior of the outgoing modes $\phi_\omega^{out}$ at $\mathscr{I}^-$. This requires tracing the modes backward in time, which introduces certain difficulties and leads to the ``transplanckian problem'' discussed in Sect.~\ref{sect:1}.

The form of the Bogoliubov coefficients in the region $v > v_0$ at $\mathscr{I}^+$ is always given by eq.~\eqref{positivemodesoutregion}. To determine their form in the Minkowski region $v < v_0$, we can impose two conditions:
\begin{enumerate}
\item Matching along the shock-wave $v = v_0$. Just before this, the form of the ``out'' modes in the $u_{in}$ coordinates reads
\begin{equation}
\phi_\omega^{out} = \frac{1}{4\pi \sqrt{\omega}}\frac{e^{-\text{i}\omega u_{out}(u_{in})}}{r}\, .
\end{equation}
We have to determine $u_{out}(u_{in})$ by matching the two metrics along the shock-wave. Requiring continuity across $v = v_0$ implies $r(v_0, u_{in}) = r(v_0, u_{out})$, where, using the relations between $(t, r)$ and $(u, v)$
\begin{equation}
\begin{split}
r(v_0, u_{in}) & = \frac{v_0 - u_{in}}{2}\, ;\\
r(v_0, u_{out}) & = \frac{v_0 - u_{out}}{2}- 2 M \ln \left[\frac{r(v_0, u_{out})}{2M}-1 \right]\, .
\end{split}
\end{equation}
Equating the above expressions gives
\begin{equation}
u_{out} = u_{in} - 4M \ln \frac{|v_0-4M-u_{in}|}{4M}\, .
\label{uoutinrelation}
\end{equation}
\item The regularity condition at $r = 0$ forces this form of the modes in the Minkowski part of the spacetime
\begin{equation}
\phi_\omega^{out} = \frac{1}{4\pi \sqrt{\omega}} \left[\frac{e^{-\text{i}\omega u_{out}(u_{in})}}{r} -\frac{e^{-\text{i}\omega u_{out}(v)}}{r} \theta(v_\text{H}-v)\right]\, ,
\end{equation}
where $v_\text{H} \equiv v_0 - 4M$ is the location of the null ray that will form the horizon at $u_{out} \to \infty$.
\end{enumerate}

We analyze the behavior of $\phi_\omega^{out}$ in the two limiting regions. At early times, i.e., $v \to -\infty$, from eq. \eqref{uoutinrelation}, we have $u_{out}(v) \simeq v$. At $\mathscr{I}^-$, therefore, we have
\begin{equation}
\phi_\omega^{out} \simeq -\frac{1}{4\pi \sqrt{\omega}} \frac{e^{-\text{i}\omega v}}{r}\, .
\end{equation}

This is still a positive-frequency mode with respect to the affine parameter $\nu$ at $\mathscr{I}^-$. Substituting this into $\beta_{\omega\omega'}$ in \eqref{betaBogoomegaomegaprime} yields zero. Consequently, at early times and asymptotically, no particle emission reaches $\mathscr{I}^+$.

As $u_{out}$ increases, instead, at late times $u_{out} \to +\infty$ (corresponding to $v \to v_\text{H}$), using eq. \eqref{uoutinrelation} we have
\begin{equation}
\phi^{out}_\omega \simeq -\frac{1}{4\pi \sqrt{\omega}} \frac{e^{-\text{i}\omega \left(v_\text{H} - 4M \ln \frac{v_\text{H}-v}{4M} \right)}}{r} \theta(v_\text{H}-v)\, .
\label{uoutlatetimes}
\end{equation}

The infinite blueshift in the exponent and the existence of the critical time $v_\text{H}$, typical of processes involving black hole formation, will be responsible for the presence of the outgoing flux at $\mathscr{I}^+$. The equation above is a superposition of positive and negative frequency modes with respect to the inertial time at $\mathscr{I}^{-}$, which is at the basis of the result of Hawking. \\
We can now provide a better intuitive argument to justify the existence of Hawking radiation. Consider an asymptotic observer in the initial region sending a light ray with very large frequency before the advanced time $\nu_{H}$. This light ray propagates through the Minkowski region without modifying its frequency $\omega$. Once it enters the Schwarzschild region, crossing the dynamical line $\nu = \nu_0$, the geometry will alter the amplitude and the frequency of the wave. The effective frequency of the outgoing modes measured by the inertial observer at $\mathscr{I}^{+}$ will suffer an exponentially increasing redshift
\begin{equation}
	\omega \sim \omega_{r} e^{-u_{out}/4M}
\end{equation}
with an e-folding time of $\kappa^{-1} = 4M$ which is the surface gravity of the Schwarzschild black hole. As a consequence of this gravitational redshift, the energy of the outgoing wave is much smaller than that of the incoming wave, without having increased the black-hole internal energy. To reestablish energy conservation, an additional emission of particles, carrying the necessary energy, is required. This should correspond to a process of stimulated emission, which in turn suggests the existence of a mechanism of spontaneous emission, corresponding to the Hawking process.

Thus, the non-stationary character of the metric is essential for the Hawking effect. Moreover, the presence of the event horizon is also crucial. Without it, the redshift is not divergent and the energy deficit in the process would be very small.

\subsection{Wave packets and computation of the Bogoliubov coefficients}
The quantity of interest is the expectation value of the number of particles with a given frequency emitted at $\mathscr{I}^+$, which, according to eq.~\eqref{expectationvalueN}, is given by
\begin{equation}
\langle in |N_\omega^{out} |in\rangle = \int_0^\infty \, \text{d}\omega' \, |\beta_{\omega \omega'}|^2\, .
\label{averageN}
\end{equation}
We assume that the quantum state of the matter fields is the natural vacuum $|in\rangle$ at $\mathscr{I}^-$. Considering the transplackian problem mentioned earlier, this assumption is quite strong. Although the matter field is in its natural vacuum at $\mathscr{I}^-$, the extreme blue-shift it experiences near the horizon could, in principle, bring it into a highly excited state. The validity of the adiabatic principle prevents this from happening (see Sect.~\ref{sect:1}).

The expression \eqref{averageN} gives the average number of created particles at $\mathscr{I}^+$ in the frequency range $\left(\omega, \, \omega+\text{d}\omega\right)$. Since $|\beta_{\omega \omega'}|$ behaves like $(\omega^{\prime})^{-1/2}$ for large $\omega^{\prime}$, this integral has a logarithmic divergence. The collapsing ball produces a steady flux of radiation at $\mathscr{I}^+$, so the infinite total number of created particles corresponds to a finite steady rate of emission continuing for an infinite time. This is due to the fact that using  states with a definite frequency implies an absolute uncertainty in time, meaning that the above expression provides the number of particles with frequency $\omega$ emitted at any time. Here, instead, we are more interested in the mean number of particles produced at late retarded time $u_{out} \to +\infty$, when the black hole has settled down to a stationary configuration. Therefore, for our purposes, it is more convenient to replace the plane-wave type modes with wave packets.
The complete orthonormal set of wave packet modes at $\mathscr{I}^+$ is the following
\begin{equation}
\phi^{out}_{jn} = \frac{1}{\sqrt{\epsilon}}\int_{j \epsilon}^{(j+1)\epsilon} \, \text{d}\omega \, e^{2\pi \text{i} \omega \frac{n}{\epsilon}} \, \phi^{out}_\omega\, ,
\end{equation}
where $n$ and $j \geq 0$ are integers, while $\epsilon$ is a parameter controlling the width of the wave packets. The latter are peaked around $u_{out} = \frac{2\pi n}{\epsilon}$, with width $2\pi/\epsilon$. Taking $\epsilon \gg 1$ ensures that the modes are narrowly centered around $\omega \simeq \omega_j = j \epsilon$. Therefore, $\langle N^{out}_{jn} \rangle$ counts the particles with frequencies in the range $(\omega_j -\epsilon, \omega_j + \epsilon)$, emitted over a time interval $2\pi/\epsilon$ at time $u_{out} = 2\pi n/\epsilon$. The late-time modes are characterized by a very large quantum number $n$, and these are the modes we must propagate backwards into the past and project onto $\mathscr{I}^-$.

To determine the number of particles created, we need to evaluate the coefficients
\begin{equation}
\beta_{jn, \omega'} = -\left(\phi_{jn}^{out}, \, \phi_\omega^{in \ast} \right) = \text{i} \int_{\mathscr{I}^-} \text{d}v \, \text{d}\Omega^2 \, r^2 \left[\phi^{out}_{jn} \partial_v \phi^{in}_{\omega'} - \phi^{in}_{\omega'} \partial_v \phi^{out}_{jn} \right]\, .
\end{equation}
Since $\phi_{jn}^{out}$ vanishes at $v \to \pm \infty$, we can perform a partial integration and discard the boundary term, yielding
\begin{equation}
\beta_{jn, \omega'} = 2\text{i} \int_{\mathscr{I}^-} \text{d}v \, \text{d}\Omega^2 \, r^2 \phi_{jn}^{out} \partial_v \phi_{\omega'}^{in}\, .
\end{equation}
Since we are primarily interested in particle production at late time, we will evaluate the above expression as $n \to \infty$. At large $n$, the wave packets at $\mathscr{I}^+$, when traced backwards in time, become localized within an infinitesimally small region around $v_\text{H}$ at $\mathscr{I}^-$. Therefore, in the limit $n \to \infty$, we can use the expression given in eq.~\eqref{uoutlatetimes}. Substituting the latter into $u_{jn}^{out}$, we obtain the following expression for $\beta_{jn, \omega'}$
\begin{equation}
\begin{split}
\beta_{jn, \omega'} & = 2\text{i} \int_{\mathscr{I}^-} \text{d}v \, \text{d}\Omega^2 r^2 \biggl\{\frac{1}{\sqrt{\epsilon}} \int_{j\epsilon}^{(j+1)\epsilon} \text{d}\omega \, e^{2\pi \text{i} \omega n/\epsilon} \times \\
&\times \biggl[-\frac{1}{4\pi \sqrt{\omega}} \frac{1}{r} e^{-\text{i}\omega \left(v_\text{H} - 4 M \ln \frac{v_\text{H}-v}{4M} \right)} \theta(v_\text{H}-v) \biggr] \partial_v \left(\frac{1}{4\pi \sqrt{\omega'}} \frac{e^{-\text{i}\omega' v}}{r} \right)\biggr\}\\
& = -\frac{1}{2\pi \sqrt{\epsilon}} \int_{-\infty}^{v_\text{H}} \text{d}v \int_{j\epsilon}^{(j+1)\epsilon} \text{d}\omega \, e^{2\pi \text{i}\omega n/\epsilon} \sqrt{\frac{\omega'}{\omega}} \, e^{-\text{i}\omega \left(v_\text{H}-4M \ln \frac{v_\text{H}-v}{4M} \right) - \text{i}\omega' v}\, ,
\end{split}
\label{betaBogocoefffinal}
\end{equation}
where in the second step, we have applied the Heaviside theta function on the extrema of the integral over $v$ and performed the integral over the angular coordinates. In the following, also the other Bogoliubov coefficient $\alpha$ will be useful. It reads
\begin{equation}
\begin{split}
\alpha_{jn, \omega'} &= (\phi^{out}_{jn}, \phi^{in}_{\omega'})\\
& = -2\text{i} \int_{\mathscr{I}^-} \text{d}v \, \text{d}\Omega^2 \, r^2 \phi^{out}_{jn} \partial_v \phi_{\omega'}^{in \ast}\\
& = -\frac{1}{2\pi \sqrt{\epsilon}} \int_{-\infty}^{v_\text{H}} \text{d}v \int_{j\epsilon}^{(j+1)\epsilon} \text{d}\omega \, e^{2\pi \text{i} \omega n/\epsilon} \, \sqrt{\frac{\omega'}{\omega}} \, e^{-\text{i}\omega \left(v_\text{H} - 4M \ln \frac{v_\text{H}-v}{4M} \right)+ \text{i}\omega' v}\, .
\end{split}
\label{alphaBogocoefffinal}
\end{equation}

We now introduce the variable $x \equiv v_\text{H} - v$. Equation \eqref{betaBogocoefffinal} becomes
\begin{equation}
\beta_{jn, \omega'} = -\frac{e^{-\text{i}\omega' v_\text{H}}}{2\pi \sqrt{\epsilon}} \int_0^\infty \text{d}x \int_{j\epsilon}^{(j+1)\epsilon} \text{d}\omega \, e^{2\pi \text{i} \omega n/\epsilon} \, \sqrt{\frac{\omega'}{\omega}} \, e^{-\text{i} \omega \left(v_\text{H} - 4M \ln \frac{x}{4M} \right) + \text{i}\omega' x}\, .
\end{equation}
The integral over $\omega$ can be readily evaluated, considering that it varies over a small interval. Since $\epsilon \ll 1$ and we are close to $v = v_\text{H}$ (so to $x \sim 0$), the dominant contributions to the integral come from the exponential term $e^{2\pi \text{i}\omega n/\epsilon + \text{i}\omega 4M \ln x/4M}$, and thus we have
\begin{equation}
\beta_{jn, \omega'} = -\frac{e^{-\text{i}(\omega_j + \omega')v_\text{H}}}{\pi \sqrt{\epsilon}}\sqrt{\frac{\omega'}{\omega_j}} \int_0^\infty \text{d}x \, e^{\text{i}\omega' x}\, \frac{\sin \left(\epsilon L(x)/2 \right)}{L(x)} \, e^{\text{i}L(x) \omega_j}\, ,
\label{betaintegral}
\end{equation}
where $\omega_j = j \epsilon \simeq (j+\frac{1}{2})\epsilon$, while
\begin{equation}
L(x) \equiv \frac{2\pi n}{\epsilon} + 4 M \ln \frac{x}{4M}\, .
\end{equation}

With the same procedure and computations, eq.~\eqref{alphaBogocoefffinal} yields
\begin{equation}
\alpha_{jn, \omega'} = -\frac{e^{-\text{i}(\omega_j - \omega')v_\text{H}}}{\pi \sqrt{\epsilon}} \sqrt{\frac{\omega'}{\omega_j}} \int_0^\infty \text{d}x \, e^{-\text{i}\omega' x} \, e^{\text{i}L(x) \omega_j} \, \frac{\sin\left(\epsilon L(x)/2 \right)}{L(x)}\, .
\label{alphaintegral}
\end{equation}
Both Bogoliubov coefficients, thus, depend on an integral of the form
\begin{equation}
I(\omega') \equiv \int_0^\infty \text{d}x \, e^{-\text{i}\omega' x} e^{\text{i} L(x) \omega_j} \, \frac{\sin \left(\epsilon L(x)/2 \right)}{L(x)}\, .
\label{integralI}
\end{equation}

From the definitions in Sect.~\ref{subsec:Bogotransfo}, we know that $\omega'>0$ for the $\alpha$ coefficients, while $\omega' < 0$ applies to the $\beta$ ones. We now proceed to derive an important relation between the two coefficients.

The integral \eqref{integralI} possesses a branch point at $x = 0$, arising from the logarithm in the definition of $L(x)$. Let us assume the branch cut to lie in the negative real semi-axis. The integrand remains analytic in the whole complex plane, except along the real negative axis. 


For $\omega'< 0$, we can rotate the contour of integration to the negative imaginary axis and then set $x \equiv -\text{i}y$. Therefore, $\ln(x/4M) = \ln(-\text{i} y/4M) = -\text{i}\frac{\pi}{2}+\ln \left(y/4M \right)$, and eq. \eqref{integralI} yields
\begin{equation}
\begin{split}
I(\omega'>0) &= -\text{i}\int_0^\infty \text{d}y \, e^{-\omega' y} \frac{\sin\left(\epsilon L_y/2 \right)}{L_y} \, e^{\text{i}L_y \omega_j}\\
L_y &\equiv \frac{2\pi n}{\epsilon} + 4M \left(-\frac{\text{i}\pi}{2}+\ln \frac{y}{4M} \right)\, .
\end{split}
\end{equation}
Similarly, if $\omega' < 0$, we can rotate the contour to positive imaginary axis and define $x \equiv \text{i}z$, from which we have
\begin{equation}
\begin{split}
I(\omega' < 0) &= \text{i}\int_0^\infty \text{d}z \, e^{\omega' z} \, \frac{\sin\left(\epsilon L_z/2 \right)}{L_z} \, e^{\text{i}L_z \omega_j}\\
L_z & \equiv \frac{2\pi n}{\epsilon} + 4M \left(\frac{\text{i}\pi}{2}+\ln \frac{z}{4M} \right)\, .
\end{split}
\end{equation}

Combining the two, we have thus
\begin{equation}
\begin{split}
I(\omega' > 0) &= -\text{i} e^{2\pi M \omega_j} e^{2\pi \text{i} n \omega_j/\epsilon} \int_0^\infty \text{d} y \, e^{-\omega' y} \, \frac{\sin\left(\epsilon L_y/2 \right)}{L_y} \, e^{4 M \text{i} \omega_j \ln(y/4M)}\, ;\\
I(\omega'<0) & = \text{i} e^{-2\pi M \omega_j} e^{2\pi \text{i} n \omega_j/\epsilon} \int_0^\infty \text{d}z \, e^{\omega'z} \, \frac{\sin\left(\epsilon L_z/2 \right)}{L_z} \, e^{4 M \text{i} \omega_j \ln(z/4M)}\, .
\end{split}
\end{equation}
For narrow wave packets ($\epsilon \ll 1$), centered around $\omega_j$, and at late times, i.e., $n \to \infty$, we have $\epsilon L_y \simeq \epsilon L_z$, and this yields the above relation between the integrals
\begin{equation}
I(\omega'>0) = -e^{4\pi M \omega_j} I(\omega'< 0)\, .
\end{equation}
Using eqns.~\eqref{betaintegral} and \eqref{alphaintegral}, this entails a relation between the Bogoliubov coefficients
\begin{equation}
|\alpha_{jn, \omega'}| = e^{4\pi M \omega_j}\, |\beta_{jn, \omega'}|\,.
\label{alphabetarelationfinal}
\end{equation}
As we shall see, this is essential in determining the spectrum of the emitted particles.

\subsection{Planck spectrum}
We can now derive the spectrum of particles detected at future infinity. The continuum version of the first equation in eq.~\eqref{Bogocoeffrelations} gives
\begin{equation}
\int_0^\infty \text{d}\omega' \, \left(\alpha_{jn, \omega'} \alpha^\ast_{j' n', \omega'} - \beta_{jn, \omega'} \beta^\ast_{j'n', \omega'} \right) = \delta_{jj'} \delta_{nn'}\, ,
\end{equation}
which, for $j = j'$ and $n = n'$ becomes
\begin{equation}
\int_0^\infty \text{d}\omega \left(|\alpha_{jn, \omega'}|^2 - |\beta_{jn, \omega'}|^2 \right) = 1 \, .
\end{equation}
Using eq.~\eqref{alphabetarelationfinal}, we also have $|\alpha_{jn, \omega'}|^2 = e^{8\pi M \omega_j}|\beta_{jn, \omega'}|^2$, which gives
\begin{equation}
\int_0^\infty \text{d}\omega' \, |\beta_{jn,\omega'}|^2 = \frac{1}{e^{8\pi M \omega_j}-1}\, .
\end{equation}
The left hand side is exactly eq.~\eqref{averageN}, namely the average number of particles detected at future infinity at late times. The right hand side coincides with a Planck distribution at a temperature
\begin{equation}
T_\text{H} = \frac{1}{8\pi M}\, ,
\end{equation}
which is the Hawking temperature.


\newpage

\end{document}